\documentclass [floatfix,  superscriptaddress, twocolumn, showpacs, aps, pra, 10pt]{revtex4-1}
\usepackage{amsmath}
\usepackage{hyperref}
\usepackage{graphicx}
\usepackage{bbm}
\usepackage{amssymb}
\usepackage[english]{babel}
\usepackage{lmodern}
\usepackage[T1]{fontenc}
\usepackage[utf8]{inputenc}
\usepackage{wasysym}
\usepackage{amssymb}
\usepackage{gensymb} %abilita il comando \degree
\usepackage{array} %abilita lo splitting del testo in tabella

\graphicspath{{SWFigs/}}

\begin{document}

\title{Validation of a fast and accurate magnetic tracker operating in the environmental field}
% Force line breaks with \\

%\newcommand{\orcidauthorA}{0000-0002-1642-5391}
%\newcommand{\orcidauthorB}{0000-0001-9091-2780}
%\newcommand{\orcidauthorC}{0000-0001-9092-3965}
%\newcommand{\orcidauthorD}{0000-0002-7298-6185}
%\newcommand{\orcidauthorE}{0000-0002-3700-2208}
%\newcommand{\orcidauthorF}{0000-0003-3495-979X}
%A vanno messi gli orcid

%\Author{Valerio Biancalana $^{1,}$*\orcidA{}, Roberto Cecchi\orcidB{} $^{2}$, Piero Chessa\orcidC{}, Giuseppe Bevilacqua$^{1}$\orcidD{}, Yordanka Dancheva$^{2,\dagger}$\orcidE{} and Antonio Vigilante $^{1,\ddagger}$\orcidF{}}

%\AuthorNames{Valerio Biancalana, Roberto Cecchi, Piero Chessa, Giuseppe Bevilacqua, Yordanka Dancheva and Antonio Vigilante}

%\address{
%$^{1}$ \quad DIISM, Siena University. Via Roma 56 53100 Siena, Italy\\
%$^{2}$ \quad DSFTA, Siena University. Via Roma 56 53100 Siena, Italy}

%\firstnote{Current address: Aerospazio Tecnologie srl, Strada di Ficaiole, 53040 Rapolano Terme (SI), Italy} 
%\secondnote{Current address: Department of Physics and Astronomy, University College London, Gower Street, London WC1E 6BT, United Kingdom}

%\corres{Correspondence: valerio.biancalana@unisi.it}

\author{Valerio Biancalana} 
\affiliation{DIISM, University of Siena -- Via Roma 56, 53100 Siena, Italy}
\email{valerio.biancalana@unisi.it}
\author{Roberto Cecchi} 
\affiliation{DSFTA, University of Siena -- Via Roma 56, 53100 Siena, Italy}
\author{Piero Chessa}
\affiliation{Department of Physics, Pisa University, Largo Pontecorvo, 3, 56127 Pisa, Italy}

\author{Marco Mandalà} 
\affiliation{DSMCN, Siena University, Viale Bracci 16  53100 Siena, Italy}
\author{Giuseppe Bevilacqua}
\affiliation{DIISM, University of Siena -- Via Roma 56, 53100 Siena, Italy}
\author{Yordanka Dancheva}
\affiliation{Aerospazio Tecnologie srl, Strada di Ficaiole, 53040 Rapolano Terme (SI), Italy}
\author{Antonio Vigilante}
\affiliation{Department of Physics and Astronomy, University College London, Gower Street, London WC1E 6BT, United Kingdom}

\begin{abstract}
{We characterize the performance of a system based on a magnetoresistor array. This instrument is developed to  map the magnetic field, and to track a dipolar magnetic source in the presence of a static homogeneous field. The  position and orientation of the magnetic source with respect to the sensor frame is retrieved together with the orientation of the frame with respect to the environmental field. A nonlinear best-fit procedure is used, and its precision, time performance, and reliability are analyzed. This analysis is performed in view of the practical application for which the system is designed that is an eye-tracking diagnostics and rehabilitative tool for medical purposes, which require high speed ($\ge 100$~Sa/s) and sub-millimetric spatial resolution.  A throughout investigation on the results makes it possible to list several observations, suggestions, and hints, which will be useful in the design of similar setups.  
}
\end{abstract}

\maketitle
%\keyword{Tracking; Magnetic tracking; Eye tracking;  Sensor Array; Eye Motion.}

%\nolinenumbers

\section*{\label{sec:introduction}Introduction}

Tracking methodologies based on magnetic field measurements have been widely studied
in the past decades and find application in several areas, including low-invasivity medical diagnostics \cite{than_ieee_12, dinatali_ieee_13}. These methodologies are based on the accurate mapping of the magnetic field generated by one or more magnetic sources, and on mathematical analysis aimed to reconstruct geometrical features (position and orientation) of those sources. Tracking apparatuses  based on permanent-magnet sources  constitute an excellent tool for wireless (and thus minimally invasive) medical applications. The availability of strongly magnetizable materials, such as Neodymium-Iron-Boron alloys, permits the generation of well detectable magnetic signals by very small size devices, with further reduction of the invasivity level.

Compared to  other state-of-the-art tracking and/or  imaging techniques based on optical detection or ultrasonic measurements,
magnetic tracking  provides an occlusion-free scheme to estimate the target position and is intrinsically responsive to its orientation. This enables a reliable localization, and accurate trajectory reconstruction, in more than three dimensions. In the simplest case of a dipolar target, five-dimensional information is normally retrieved, consisting in (3D) position and (2D) angular orientation: the system has a blind angular component, which corresponds to rotations around the dipole direction.

The problem of retrieving position and orientation of a magnetic dipole has been successfully faced both on the basis of linear, deterministic methods \cite{meng_ieee_07} and non-linear best-fit approaches \cite{weitschies_ieee_94, schlageter_saa_01}, as well as with a combination of the two, where the linear solution is used as an input-guess for the non-linear calculation \cite{meng_ieee_08}. 

At the expense of a heavier computation, the non-linear best fit approach permits more accurate evaluation, also thanks to the possibility of using larger data sets. Its main drawback is constituted by the need of providing an appropriate initial guess to the numerical algorithm, i.e. on a limited reliability caused by the  non-convexity of the strongly non-linear function to be minimized. 
The typical working principle of best-fit procedures is based on finding a path in the parameter space, which efficiently brings from an initial, arbitrarily assigned guess to the best fitting parameter set. The latter is identified as the one that  minimizes the difference between measured quantities and their theoretical values. 
The assignment of an appropriate initial guess has a twofold relevance. It helps accelerating the convergence of the  algorithm, and --more important-- helps preventing that the algorithms converge to local minima, i.e. to  wrong  solutions. Unfortunately, the mathematical complexity which renders these numerical approaches a favorite choice, makes also difficult to elaborate reliable predictions of the guess appropriateness. Empiric tests are necessary to quantify the robustness of the best-fit procedure with respect to the accuracy level of the initial guess. 

The non-linear approach can be easily extended to more complex problems, as in the cases of multi-target tracking \cite{meng_ieee_16} or when operating in the presence of a background field, which is the case considered here. The latter feature characterizes the device described in this work, because of two main reasons: firstly it makes possible to operate with small targets in the presence of Earth field without suffering of tracking distortions induced by the latter, secondly it allows the simultaneous tracking of the magnetic target with respect to the sensors and of the sensors with respect to the ambient field.
This  is of interest in diagnostic measurements focused on the correlation of head and eye motion, which, with instrumentation based on  alternative kind of measurements, requires the fusion of data recorded by detectors of different nature \cite{dai_ieee_18, woehle_sens_20}. 

Up to date the gold standard for eye-tracking system in medical field is represented by magnetic search coil systems for research purposes and infrared head mounted cameras for clinical use. Typical angular and temporal resolution of the two different techniques are respectively $<0.1\degree$ and $>1000$~Hz (magnetic search coil) versus $0.5\degree$ and  $100-200$~Hz (high-speed,  head mounted infrared cameras) \cite{Agrawal_oen_14}. 
The last option is the most widely used in the medical field since it is less expensive, non-invasive and portable. Despite of that, infrared cameras have several drawbacks in terms of eye movement recording: lower spatiotemporal resolution, difficulties in recording vertical and torsional eye movements, difficulties with light eyes, artifacts due to blinks, need the camera to be strongly tightened to the head, need of data fusion to record also head movements, heavy computational burden for image analysis. An overview about speed and angular resolution of low-cost eye trackers developed in the last decade can be found in ref.\cite{rakhmatulin_arx_20}.

The paper is organized as follows: Sec.\ref{sec:sensor} provides a brief description of the hardware setup and of its specifications; Sec.\ref{sec:model} describes the methodology applied to extract tracking information from magnetometric data; Sec.\ref{sec:trajectory} reports a set of tracking outputs and estimates of tracking accuracy; Sec.\ref{sec:timeperformance} describes an analysis devoted to quantify the achieved speed under different operation conditions of the numerical algorithm application; Sec.\ref{sec:2dmaps} deepens the analysis of the time performance and studies the relevance of providing an appropriate initial guess to the numerical algorithm. Finally, a synthesis of the achievements and potentialities is provided in the following two sections.

\section{Setup}
\label{sec:sensor}
\begin{figure}[ht]
   \centering
        \includegraphics [angle=00, width=  \columnwidth] {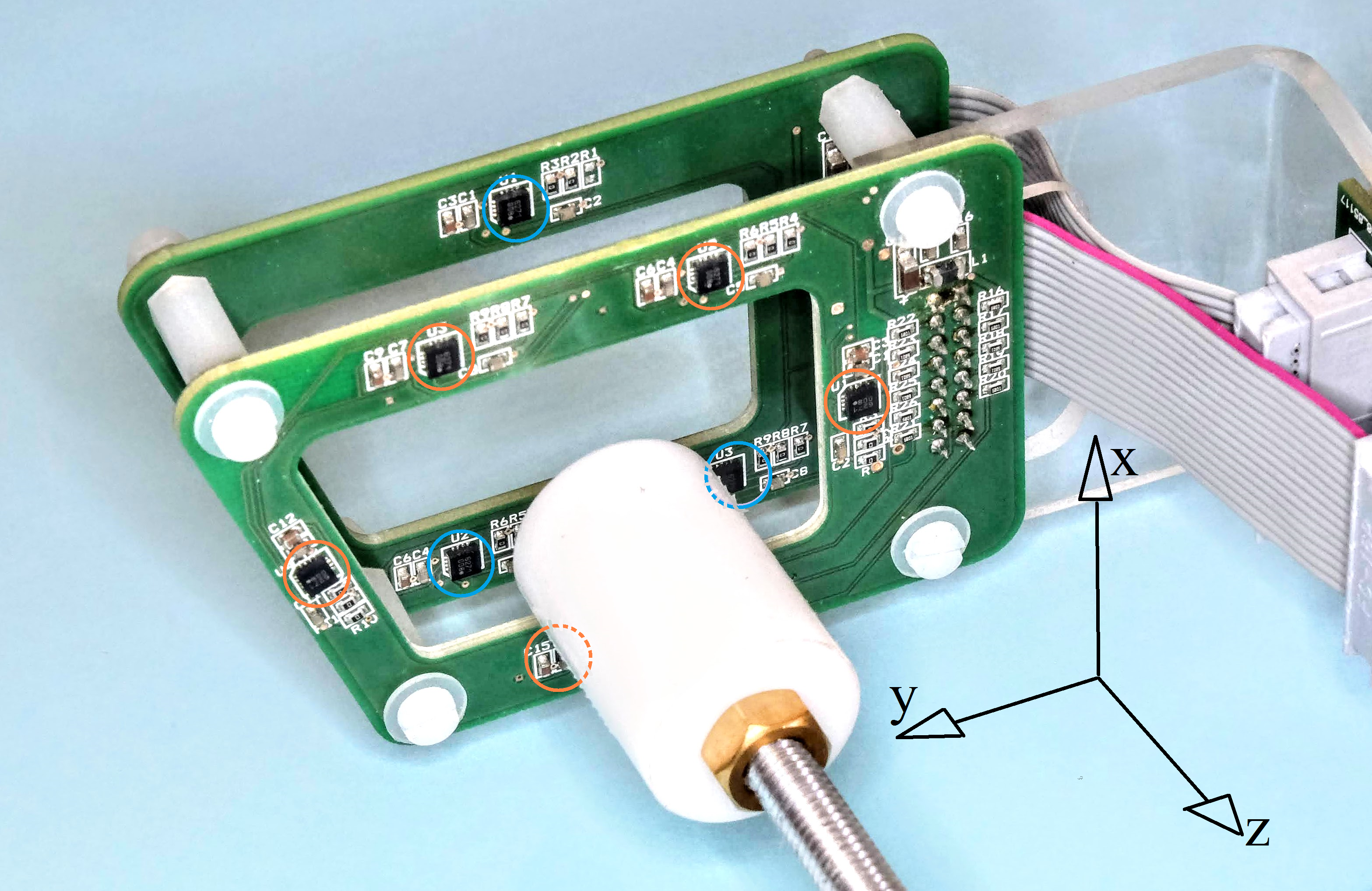}
        \caption{The sensor array contains 8 sensors distributed on two parallel PCBs (three of them, highlighted with blue circles are a $z=0$ and other five  (orange circles) at $z=16.6$~mm). The system is designed for eye-tracking purposes. A 20~mm diameter cylinder is used to emulate eye motion. It hosts the magnetic source (target), which  performs either circular or helix trajectories.  }
  \label{fig:sensorarray}
\end{figure}
The hardware of the setup is extensively described in the Ref.\cite{biancalana_instr_21}. Briefly, eight triaxial magnetoresistive sensor chips (Isentek IST 8308) are  mounted on two parallel, rigidly assembled printed circuit boards (PCBs). The chips  are driven by a microcontroller, which synchronizes their acquisition and makes possible to memorize their readings (for calibration purpose) or to transmit them in real time to a computer, via USB interface (for data storage and/or online tracking).
Beside the three sensors, each chip contains a triple 14~bit, 200~Sa/s analog-to-digital converter and numerical filters for noise reduction. Prior to normal (tracking) operation, calibration  data are initially collected in a uniform field and analyzed to define a set of conversion coefficients saved into the computer. These conversion coefficients are used during the normal operation to:
\begin{itemize}
    \item Subtract the individual sensor offsets;
    \item Make the responses isotropic;
    \item Refer to a unique co-ordinate system, oriented in accordance with that of a sensor selected as the reference one;
\end{itemize}
that is, in synthesis, to convert the raw readings in accurate and consistent magnetometric data.

During the tracking measurements, the resulting homogeneous, calibrated and equalized magnetometric data are numerically analyzed by means of a Levemberg-Marquardt algorithm \cite{levenberg_qam_44, marquardt_jsiam_63} to localize the magnetic source and to determine vectorially the environmental field and the magnetic dipole of the target.

The  permanent magnets used in this experiment are disks 2~mm in diameter and 0.5~mm in thickness. Their volume --considering the typical magnetization of Neodymium-Iron-Boron materials ($M\approx 1$~T/$\mu_0$)-- gives a magnetic dipole $m\approx 10^{-3}$~Am$^2$. Such a dipole produces a magnetic field comparable with the Earth one ($\approx 40\, \mu$T) at centimetric distances. The tests presented in this work are performed with the magnet attached on the surface of the white cylinder visible in Figure~\ref{fig:sensorarray}. The cylinder rotates around its axis which can be alternatively plain or threaded, as to produce circular or helix trajectories.

\section{Model and algorithms}
\label{sec:model}
Each measurement produces $3K$ ($K=8$) magnetometric data corresponding to the three components of the magnetic field detected in the (nominally known) positions $\vec r_k$ of the $K$ sensors.

\subsection{Field model}
\label{sec:model:fieldmodel}
The field measured by the $k$th sensor when the dipole $\vec m$ is localized in $\vec r$ is modeled as:

\begin{equation}
\vec B_k (\vec r, \vec m) = \frac{\mu_0}{4 \pi}\left ( 3\frac{[\vec m \cdot (\vec r_k-\vec r)](\vec r_k-\vec r) }{|\vec r_k-\vec r|^5}
-\frac{\vec m}{|\vec r_k-\vec r|^3}
\right)+\vec B_g,
\label{eq:modellodipolo}
\end{equation}
where $\vec r_k$ ($k=0, \dots, K-1$) is the position of the $k$th sensor  and $\vec B_g$ is the environmental field, which is assumed to be homogeneous over the sensor-array volume.

Despite the highly non linear $\vec B_k (\vec r, \vec m)$ dependence, the inverse problem of determining $\vec r$ and $\vec m$ from a minimal set of $\vec B_k (\vec r)$ measurements performed in known $\vec r_k$ positions has been successfully approached, under conditions of negligible $\vec B_g$ \cite{meng_ieee_07}.

However, also in those conditions, a non-linear best fit approach, making use of larger data sets from many sensors, helps reduce noise and imperfection effects, and produce better tracking results at the expense of a heavier computation burden.
The latter problem is efficiently circumvented by the currently available computers, and its main drawback actually consists in the risk that best-fit convergence conditions are not fulfilled.

\subsection{Limits}
\label{sec:model:limits}
\begin{figure}[ht]
   \centering
        \includegraphics [angle=00, width= 0.9 \columnwidth] {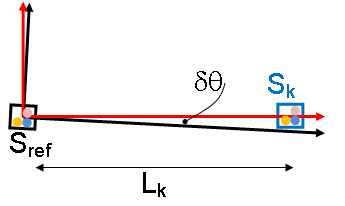}
        \caption{The field components are referred to co-ordinates defined within the reference sensor chip $S_{ref}$ (in black), which might be slightly misaligned with respect to the PCB frame (in red). This can lead to inaccuracy in the position of the other chips ($S_k$). In addition, each chip contains three sensors (represented by colored dots), which are (submillimetrically) displaced with respect to each other.   }
  \label{fig:misalign}
\end{figure}
One of the possible imperfections is related to the limited precision with which the sensor positions are known. Firstly, each chip contains three separate magnetoresistive sensors (represented with dots in the Figure~\ref{fig:misalign}) for the measurement of the three field components, and they are not located in one point within the chip. This limit could be (at least partially) mitigated assigning the precise locations; however the relative displacements are sub-millimetric and do not play a substantial role, unless the system to be built is scaled down to sub-centimetric size.
Secondly, there could be a misalignment between the co-ordinate axes within the reference sensor (that used to define the field components) and the PCB reference frame (that used to define the positions of the other sensors with respect to the reference one). As schematically shown in Figure~\ref{fig:misalign}, an angular misalignment $\delta \theta$ would produce systematic errors $L_k \delta \theta$ in the determination of the $k$th sensor position, $L_k=|\vec r_k - \vec r_0|$ being its distance from the reference one.  The problem is 3-D and three angular uncertainties concur to determine the sensor position errors. In our implementation the $L_k$ are in the cm scale, and $\delta \theta$ might be of the order of one-degree, which leads --again-- to submillimetric errors. Additional (but even smaller) uncertainties are related to minor chip translations with respect to their ideal positions on the PCB. 

Best-fit procedures (based on data consistency) may help determine the sensor displacements with respect to their nominal positions. We have  successfully attempted to face the problem with this approach, obtaining only submillimetric corrections and barely appreciable improvements in terms of tracking accuracy: this activity  will not be discussed further in this paper.

Another assumption made to derive eq.\ref{eq:modellodipolo} is the punctual nature of the source, which is indeed modeled as a pure dipole. This assumption is very well justified by the small size of the magnetic source ($\le 1$~mm) with respect to the source-sensors distance ($> 1$~cm): higher order multipole terms contribute to the field in a definitely negligible manner.

When strong magnets are used, the environmental field can be neglected, and in this case its presence is regarded as an external interference which may deteriorate the tracker performance. 
In contrast, we use of relatively weak magnetic sources, whose size is selected in such a way to make the ambient field and the dipole field on the sensors of comparable intensities. Under this condition, the ambient field is far from being negligible, while it can be well identified by the best-fit procedure.
An alternative approach that could be used when a non-negligible ambient field is present, is to detect it with a far-located sensor (on which it is the dipole field that is negligible) and to subtract it to the other measurements prior to the data elaboration.
At the expense of an increased number of fit parameters, our approach brings a threefold advantage: it permits a more accurate evaluation of the ambient field (data from all the sensors  concur to its determination), it makes the system less affected by ambient field inhomogeneities (the gradiometric baseline is set by the array size), the size of the sensor array is maintained small (it does not contain a far located reference sensor).
As already verified (Figure~6 in Ref.\cite{biancalana_instr_21}), the target tracking results substantially unaffected by $B_g$ variations, which demonstrates the feasibility and reliability of this approach. 

\subsection{Best-fit degrees-of-freedom}
\label{sec:model:dof}
Referring to eq.\ref{eq:modellodipolo}, the best fit procedure is in charge of retrieving $n_p=9$ parameters, and namely the components of the 3D vectors $\vec r, \vec B_g, \vec m$. In principle, being the intensity of $\vec B_g$ and $\vec m$ assigned, a reduced fitting parameter set could be used. In our experiments we had the evidence that setting a fixed $B_g$ modulus severely worsens  the performance (accuracy and reliability), as soon as its real value changes because of sensors  displacements. The reason for that is the typical presence of smooth but non-negligible ambient field inhomogeneities caused, e.g. by ferromagnetic materials of buildings and furniture. In contrast, it can be definitely advantageous using a fixed $| \vec m|$, as to reduce the best fit to determine only the dipole orientation. Provided that $m$ is accurately evaluated, this reduction of $n_p$ from 9 down to 8 improves the tracking performance, as discussed below in Sec.\ref{sec:trajectory:helix}. Differing from what one could expect, passing from a 9-parameter procedure (9P) to an 8-parameter one (8P), does not help reduce the number of iterations or shorten the computation time, as discussed in Sec.\ref{sec:timeperformance}.

The Levemberg-Marquardt algorithm used for the analyses here presented has a termination condition based on the comparison of a tolerance parameter ($T$) with the variation of a mean residual $R$, which is defined as the root of the mean squared deviation between the estimated and the measured values of the fields. The fitting procedure is terminated and a parameter estimation is output as soon as an iteration produces a relative variation $\Delta R / R < T$. All the results reported in this paper are obtained with $T$ set at $10^{-6}$. However, we verified that selecting a 100 times tighter condition ($T= 10^{-8}$) increases the  number of necessary iterations by a few units (typically less than 10), both in the 8P and in the 9P case.

\subsection{Scalability}
The equation \ref{eq:modellodipolo} inherently expresses an important scale-law. For a given material remanence, the dipole intensity $m$ scales with the volume of the magnet, i.e. with the third power of the magnet size. For a given spatial positioning (location and orientation) the dipole field scales with a $1/r^3$ law, while $B_g$ is obviously not dependent on the system size. In conclusion, the same measurement results will be obtained if the whole system is stretched or shrunk by a given scale factor. Of course this scale-law would fail when the shrinking factor would be so large to make the punctual approximation of the sensor size inappropriate, while stretching can result excessive whenever the increase of the array size  causes $B_g$ inhomogeneities play a relevant role (in other terms, the mentioned scale-law would persist only if the distance of ferromagnetic disturbers was stretched as well).

\section{Trajectory Performance}
\label{sec:trajectory}
The precision with which position and trajectories can be retrieved depends on diverse parameters, among which:
\begin{itemize}
    \item accuracy of the calibration and subsequent consistency of the magnetometric data
    \item accuracy of knowledge of the sensor positions
    \item presence of time-dependent noise or other disturbances preventing accurate measurements
    \item presence of relevant inhomogeneities of the ambient field.
\end{itemize}
In addition, the condition of single measurement may greatly affect the tracking accuracy, in particular, a crucial role is played by:
\begin{itemize}
    \item intensity of the target dipole $|\vec m|$;
    \item distance of the target from the sensor array
    \item position and orientation of the target with respect the sensor array (the performance is not isotropic around the array center, and the anisotropy depends also on the dipole orientation)
\end{itemize}
In contrast, the initial guess has no effects on the tracking precision, provided that it is suitably assigned to guarantee correct convergence, that is to let the best-fit procedure get the global minimimum.  A target trajectory is reconstructed and sampled as a sequence of  tracked points. The trajectories reported in this section are obtained using a default guess (see Tab.\ref{tab:default}) to determine the first point, and then assigning each tracked point as an initial guess for the subsequent one. The latter detail has relevance for the computation time (see Secs.\ref{sec:timeperformance} and \ref{sec:2dmaps}), but not for the tracking accuracy and precision. 

\begin{table}
\centering
%% \tablesize{} %% You can specify the fontsize here, e.g., \tablesize{\footnotesize}. If commented out \small will be used.
%\tablesize{\footnotesize}
\begin{tabular}{cccc}
%\toprule
\textbf{quantity} & \textbf{x }	& \textbf{y }	& \textbf{z }\\
%\midrule
$\vec r$ & 20~mm & $-20$~mm & 40~mm \\ 
 \hline
$\vec B_g$ & 20~$\mu$T & 20~$\mu$T & 20~$\mu$T\\
 \hline
$\vec m$ & 600~$\mu$Am$^2$ & 600~$\mu$Am$^2$ & 600~$\mu$Am$^2$ \\
%\bottomrule
\end{tabular}
\caption{Default values used as starting guesses (unless differently specified) for the target position $\vec r$, the environmental field $\vec B_g$, and the magnetic dipole $\vec m$.}
\label{tab:default}
\end{table}

A general and throughout study of the possible scenarios is not feasible, while having in mind a specific use lets reduce the number of possible configurations and makes a focused characterization possible.

Our system is designed for eye-tracking and, considering the size of the "lenses like" frame shown in Figure~\ref{fig:sensorarray}, we are interested in cases where the target is located on a given side of the array (in particular $z>z_2$, having set $z_1=0$ for the three-sensor PCM and $z_2=16.6$~mm for the five-sensor one), while $(x,y)$ will not go more than some tens of mm away from the sensor axis. The white cylinder in Figure~\ref{fig:sensorarray} constitutes a reasonable representation of that region of interest (RoI).

Any quantification of the system performance, will depend on the specific positions/trajectories selected. On the other hand, any estimate performed with the target in the mentioned RoI will be a significant example, such to provide a  performance evaluation. 
In this section, we are reporting a set of such significant examples, providing quantitative or qualitative (graphical) information about the precision achieved.

\subsection{Circle}
\label{sec:trajectory:circle}
\begin{figure*}[ht]
   \centering
        \includegraphics [angle=00, width= 0.8 \textwidth] {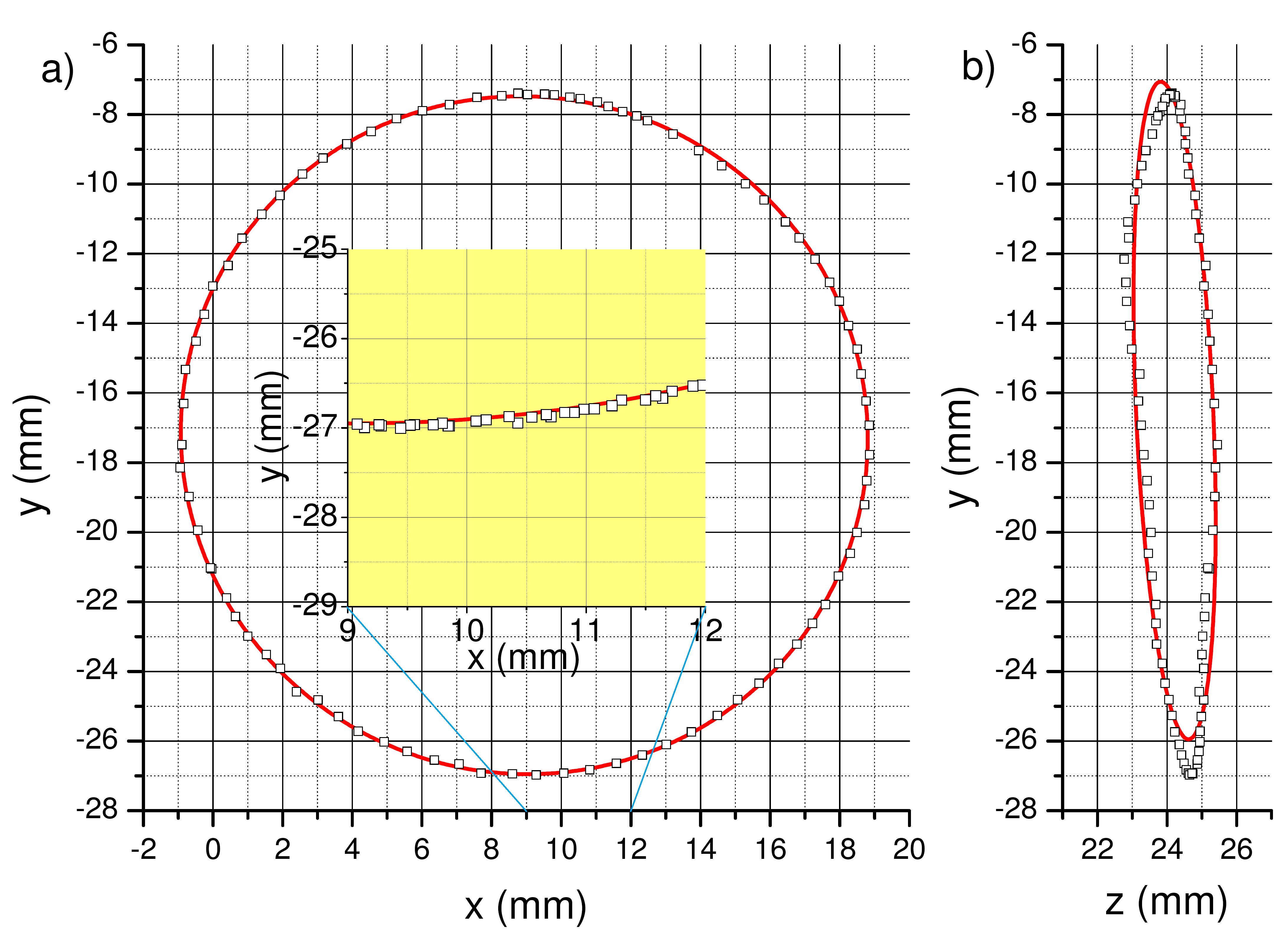}
        \caption{Projections on the $xy$ e $zy$ planes of a 8P reconstructed circular trajectory. The red curves are elliptical best-fits. Residuals as small al 6.1~$\mu$m and $92\mu$m are obtained, respectively. The corresponding 9P analysis lead to  results that are similar (5.6~$\mu$m) in the $xy$ projection  and  worse (95~$\mu$m) in the $xy$ projection.
        }
  \label{fig:cerchi}
\end{figure*}
We present a test performed by imposing a circular trajectory, on a plane nearly parallel to the PCBs.

The Figure~\ref{fig:cerchi} shows its $xy$ and $zy$ projections, together with  best-fitting curves based on an elliptical model. A 8P tracking is applied, where $m=| \vec m|$ is previously estimated as the mean value on the whole trajectory reconstructed by a 9P one ($\langle m \rangle=(995\pm60)~\mu$Am$^2$). The 9P result (not reported) is indistinguishable in the $xy$ projection and worse in the $zy$ one. The latter is a typical behavior, and there is evidence that a less accurate tracking is obtained --particularly with 9P-- along the direction perpendicular to the dipole ($z$ in the present case, the dipole being radially oriented).
The elliptic fit enables an estimate of the major axis (19.9mm), which matches the cylinder diameter within a 0.1~mm accuracy.

The data analysis that produces these spatial trackings let also determine the dipole components and quantify  the uncertainty $\delta \theta$ affecting the angular estimate of the dipole orientation. The measurement and data elaboration performed to this end are as follows. The angle describing the dipole orientation around the rotation axis (which is nearly parallel to $z$) is estimated as $\theta= \arctan(m_y/m_x)$. After a phase unwrapping procedure, the angles $\theta_n=\theta(t_n)$ are modeled and fitted with a linear function, which correspond to consider a constant angular velocity $\omega$ of the magnet driver. The residual of this linear regression provides the mean squared deviation between the estimated ($\omega t_n+\theta_0$) and the measured $\theta_n$ values of the dipole orientation. This evaluation gives   $\delta \theta_{8P}= 0.93\degree$ and $\delta \theta_{9P}= 0.78\degree$, with the 8P and 9P algorithms, respectively. It is worth mentioning that in the eye tracking application the eye orientation can  be retrieved both directly from the orientation of $\vec m$ and indirectly from $\vec r$,  assuming that the magnet is constrained to move on a spherical surface. This redundancy could be used to improve the angular resolution.

\subsection{Helix}
\label{sec:trajectory:helix}
\begin{figure*}[ht]
   \centering
        \includegraphics [angle=00, width= 0.49 \textwidth] {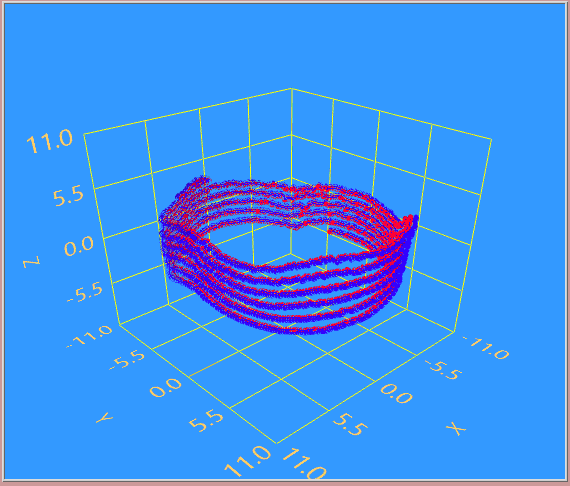}
        \includegraphics [angle=00, width= 0.49 \textwidth] {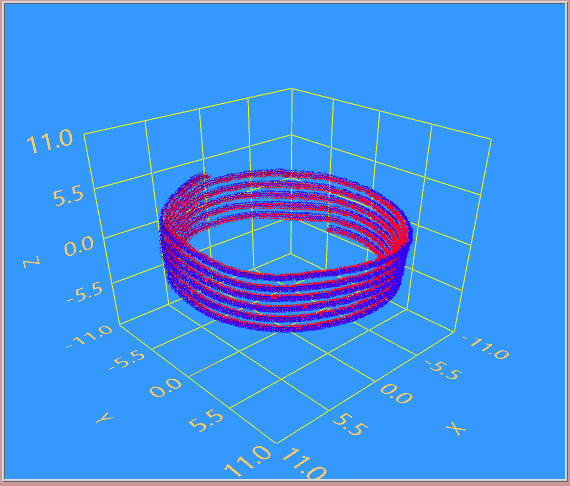}
        \caption{3D views of reconstructed helix trajectory, the axes units are expressed in mm. The magnet is radially oriented and follows a helix trajectory 10~mm in radius and 1~mm in pitch, around the $z$ axis. Beside the spatial position the tracker retrieves the dipole vector: red-blue dots are used to represent this data. The left tracking is obtained by 9P and suffers of evident (nevertheless submillimetric) distortions of the $z$ co-ordinate. The right tracking uses the same magnetometric data, but a 8P analysis.}
  \label{fig:3dhelix}
\end{figure*}
A more complex trajectory with 3D reconstruction permits performance evaluations that are qualitative, but --nevertheless-- significant. To this aim, we apply 3D trajectories to the target magnet. In this case, the magnet holder is mounted on a screw-axis, with a 1~mm pitch thread. The trajectory has consequently an helix shape, 20~mm in diameter and ~1mm in pitch. The sample 3D  trackings shown in the Figure~\ref{fig:3dhelix} represent the reconstructed trajectories obtained on the basis of a 9P and 8P, respectively. Also in this case, the dipole is radially oriented. The 9P produces evident distortions,  particularly due to undesired fluctuations of the $z$ co-ordinate. Remarkably, these $z$ distortions occur simultaneously with unexpected variations of the $m$ estimates. This feature is likely due to the  imperfections mentioned in the Sec.\ref{sec:model:limits}. The 8P greatly helps   improve the reconstructed trajectory in this respect.

\begin{figure}[ht]
   \centering
        \includegraphics [angle=00, width= 0.7 \columnwidth] {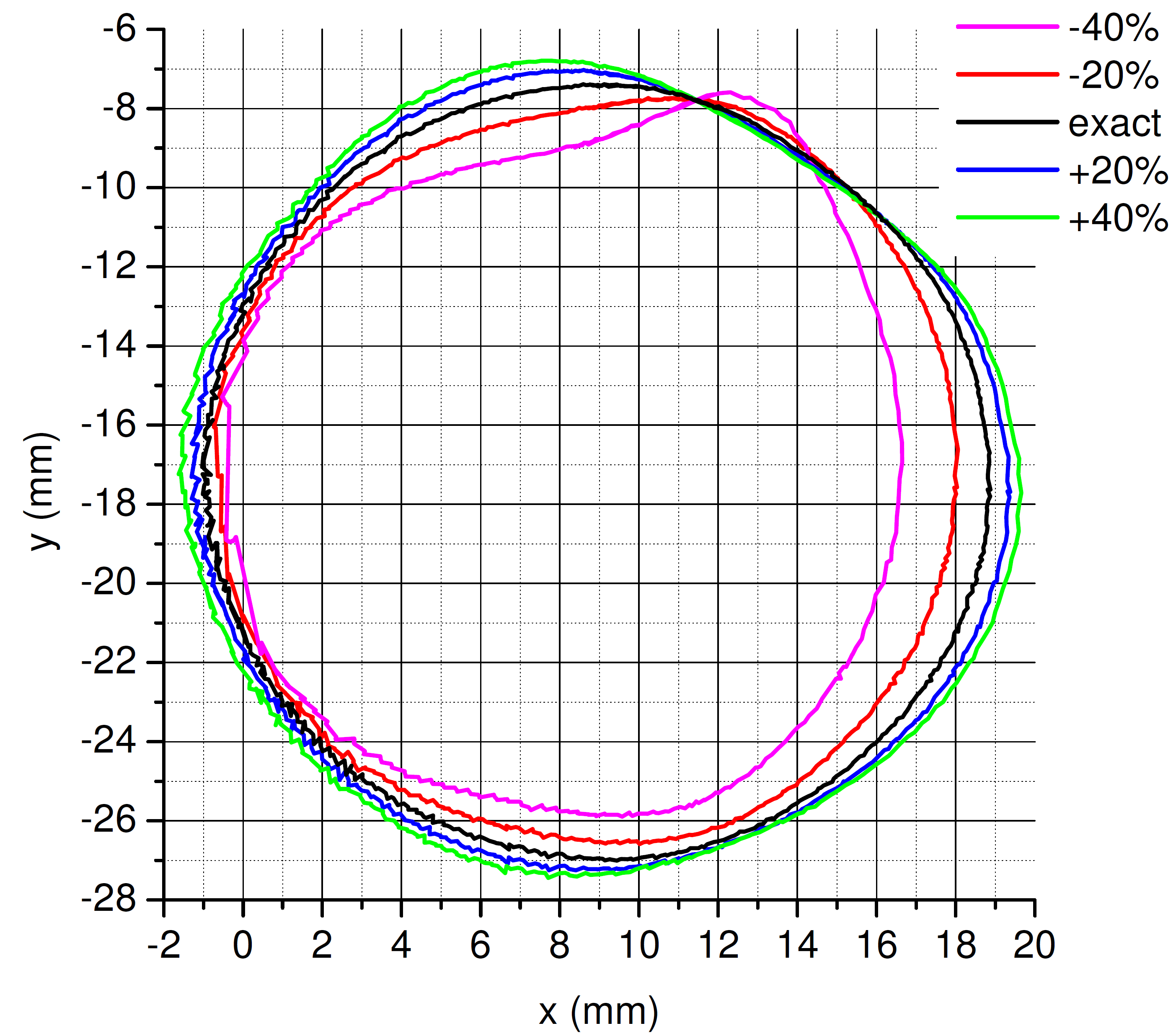}
        \caption{Tracking of $xy$ projections as obtained with 8P, using for $|\vec m|$ its value resulting from 9P (black)  and values increased (blue, green) or decreased (red, magenta) by 20\% and 40\%, respectively. }
  \label{fig:cerchistorti}
\end{figure}
\subsection{Tuning the 8-parameter fit}
The better performance of the 8P  comes at the expense of a computationally heavier model formula (it includes several trigonometric functions, which partially compensates the reduction of the parameter space dimensions) and the need of providing the procedure with an accurate estimate of  the modulus $| \vec m|$. To this end, it is possible to start with a 9P evaluation of the mean value of $m$ over a trajectory, and subsequently use that estimate of $m$ for 8P tracking. The Figure~\ref{fig:cerchistorti} shows the effects of providing 8P with wrong estimations of $m$. We  report $xy$ projections of five trackings obtained with correct (9P estimated), overestimated and underestimated $m$ values, respectively: evident trajectory distortions occur, which go rapidly in excess of 1 mm.

\section{Time performance}
\label{sec:timeperformance}
The algorithm speed becomes a critical feature whenever a  real-time output or large data-sets analysis are required. Assigning an appropriate guess not only plays a role in guaranteeing convergence to the right solution, but helps reduce the convergence time. 
\begin{figure}[ht]
   \centering
        \includegraphics [angle=00, width= 0.7 \columnwidth] {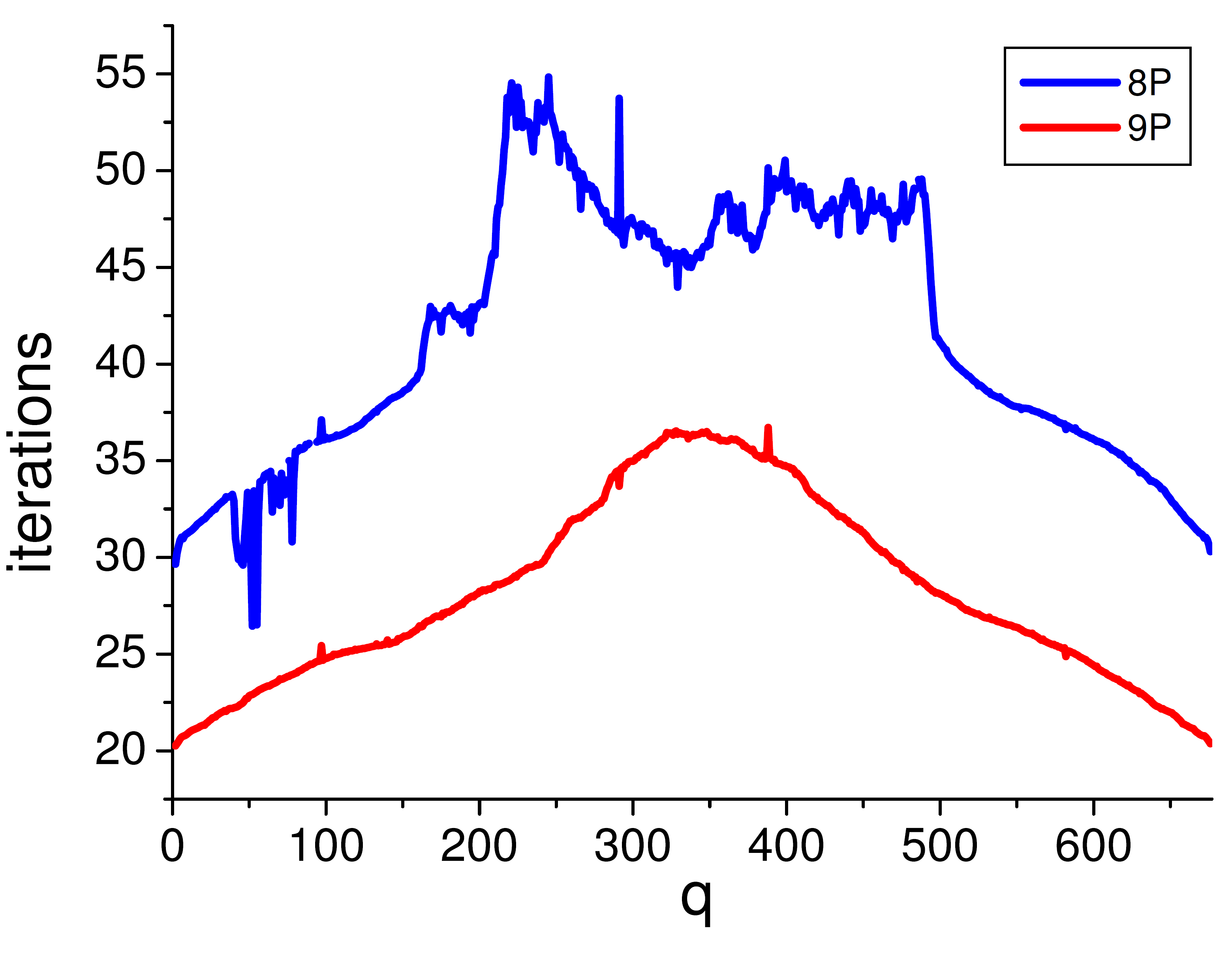}
        \caption{Average iteration number as a function of the guess accuracy for the 8P (blue) and 9P (red) cases. The cases with q=1 and q=670 correspond to using one of the nearest neighbors as an initial guess, while q=335 corresponds to using diametrically opposite points.     }
  \label{fig:convergencetime}
\end{figure}
The Figure~\ref{fig:convergencetime} provides an example of time-performance estimation as a function of the guess accuracy. The plots represent the algorithm iteration number necessary to track points of a circular trajectory (same data as for Figure~\ref{fig:cerchi}) sampled in $N=670$ points, as a function of the guess accuracy. To this end, for the $j$th tracking point, we used as starting guess the tracking output $j-q$  (or $N+j-q$, when $q>j$), with $q$ ranging from 1 (guess coming from the previous point in the tracking list) to $N-1$ (guess from the following point). 
In other terms, setting $q=N/2$ corresponds to assign as a guess the diametrically opposite point, that is a 20~mm displaced position and reversed dipole orientation. 
 Using both 8P and 9P, a correct convergence occurs in all the cases, and for both  8P and 9P the average iteration number increases by about 50\% in the worst case ($q=N/2$). 9P requires less iterations than 8P (20-35 instead of 30-55), for comparison, the average iteration number is 48 (for 8P) and 25 (for 9P) when  default guesses are assigned. 
The number of iterations $n_i$ is approximately proportional to the computation burden. The number of function calls $n_c$ depends on $n_i$ according to $n_c=(n_i+2)(2 n_p+1)$. The number $n_c$ is the quantity that, together with the complexity of the model formula, mainly determines the best-fit computing time. The  conversion factor from calls to time is machine-dependent. In our case, with an Intel-I7 2.4~GHz CPU we obtain about 14 and 13~$\mu$s/call for the 8P and 9P respectively. The consequent time performance is summarized in Tab.\ref{tab:timeperformance}, which (with the exception of the 8P worst cases) shows that both 8P and 9P require less than 10ms/tracking, as to enable a real-time 100~Sa/s acquisition.
\begin{table}%[H]
\centering
%% \tablesize{} %% You can specify the fontsize here, e.g., \tablesize{\footnotesize}. If commented out \small will be used.
%\tablesize{\footnotesize}
\begin{tabular}{cccc}
%\toprule
\textbf{quantity} & \textbf{q=1}	& \textbf{q=N/2}	& \textbf{default}\\
%\midrule
8P (iter. number)& 30 & 55 & 48 \\ 
 \hline
9P (iter. number)& 20 & 35 & 25\\
 \hline
8P (ms/tracking) & 7.5 & 13.8 & 12\\
 \hline
9P (ms/tracking) & 5.4 & 9.5 & 6.8\\
%\bottomrule
\end{tabular}
\caption{Time performances of 8P and 9P, in the case of next-neighbor guess, in case of diametrically opposite guess and in case of default values assigned programmatically (see Tab-\ref{tab:default})}
\label{tab:timeperformance}
\end{table}
Further analyses of the convergence speed are provided at the end of next section, after some considerations about the conditions under which a correct convergence is obtained.

\section{Guess criticality}
\label{sec:2dmaps}
In this section we study the criticality of providing an appropriate initial guess to make the algorithm converge to the correct solution and we give some additional information about the convergence speed.

For each given configuration ($\vec r, \vec B_g, \vec m$) of the target, there exists a volume surrounding that configuration, which corresponds to guesses that will guarantee a correct convergence. Guesses chosen out of that volume will produce wrong solutions. We have verified that the convergence condition is weakly affected by the initial values assigned to $\vec B_g$ and $\vec m$, while the good-convergence volume $V_G$ in the subspace of the geometrical  $x, y, z$ co-ordinates presents a large variety of shapes.

Intersecting of the guess volumes with planes parallel to the co-ordinate planes $xy$, $xz$ or $yz$ provides a 2D visualization of sections of the convergence volume $V_G$ and of its complement $V_R$, corresponding to the spatial guesses that lead to wrong convergence.

The figures here presented are 2D maps obtained using 9P to identify the mentioned sections: two spatial parameters of the initial guess are varied while the remaining 7 are kept at fixed values. The good-guess regions $V_G$ are represented in green and the bad-guess regions $V_R$ are represented in red. A color scale is used to represent the iteration numbers: bright colors denote fast convergence and dark colors denote slow convergence (the iteration numbers and the corresponding color scales are reported in the bottom rules of each map). The target position is represented by a white dot, and the sensors are represented in colors: blue dots for the three on the $z=0$ PCB and orange dots for the five ones on the $z=16.6$~mm PCB. All the maps refer to 300~mm~$\times$~300~mm areas on the intersecting planes, such to be significantly larger than the RoI for the eye-tracking application. The  table \ref{tab:positions} reports the co-ordinates of the sensors and of two analyzed positions of the target, both selected in our RoI, the first one in the vicinity of the sensor array axis and the second one in a more peripheral location.

\begin{table}%[H]
\centering
%% \tablesize{} %% You can specify the fontsize here, e.g., \tablesize{\footnotesize}. If commented out \small will be used.
%\tablesize{\footnotesize}
\begin{tabular}{cccc}
%\toprule
\textbf{object} & \textbf{x (mm)}	& \textbf{y (mm)}	& \textbf{z (mm)}\\
%\midrule
sensor 0& 0 & 0 & 0 \\ 
 \hline
sensor 1& 0 & -27.05 & 0\\
 \hline
sensor 2& 40.64 & -19.28 & 0\\
 \hline
sensor 3& 0 & -19.28 & 16.6\\
 \hline
sensor 4& 20.26 & -45.74 & 16.6\\  
 \hline
sensor 5& 40.64 & -32 & 16.6 \\
 \hline
sensor 6& 40.64 & -6.6 & 16.6\\ 
 \hline
sensor 7& 20.33 & 7.25 & 16.6\\
 \hline
$\vec r_c$ & 13 & -20 & 30.5\\
 \hline
$\vec r_p$ & 33 & -19.8 & 29.7\\
 \hline
%\bottomrule
\end{tabular}
\caption{Co-ordinates of the sensors. The last two lines report the central and peripheral target positions (determined by the tracker) considered in the analyses of this Section.}
\label{tab:positions}
\end{table}

\begin{figure*}[h]
   \centering
        (a)\includegraphics [width= 0.45 \textwidth] {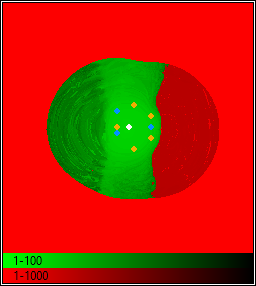}
        (b)\includegraphics [width= 0.45 \textwidth] {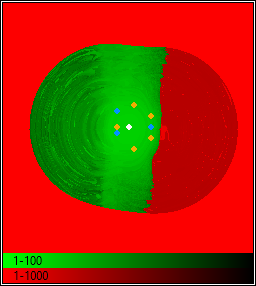}
        (c)\includegraphics [width= 0.45 \textwidth] {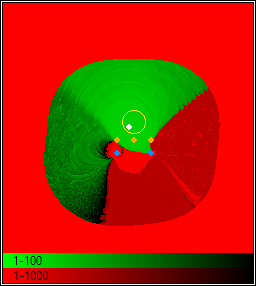}  
        (d)\includegraphics [width= 0.45 \textwidth] {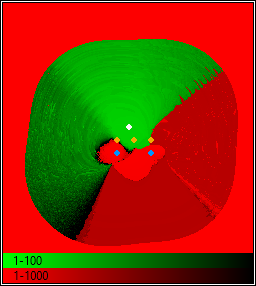}         
        
        \caption{2D convergence maps of $xy$ (a,b) and $xz$ (c,d) sections. The remaining 7 parameters are set to their exact values (a,c) or to their default values (b,d). The target (white dot) is in a nearly axial position, at $z=31$~mm. Unexpectedly, using wrong guesses for $\vec B_g$ and $\vec m$ and wrong (larger) $z$ makes the ($x,y$) convergence area larger than using exact values. All these maps describe a 300~mm$\times$300~mm area, the yellow circle in (c) corresponds to a sphere (26~mm in diameter) that is the typical size of human eye bulb.  }
  \label{fig:2dmapcentre}
\end{figure*}

The Figure~\ref{fig:2dmapcentre} represents convergence maps obtained when the dipole is located in $\vec r_c$, i.e. proximally to the sensor axis. The four maps correspond to $xy$ and $xz$ sections, obtained when assigning the fixed 7 parameters with their correct values or with default values (details provided in the caption). The sections of $V_G$ are quite regular and symmetric. There is a convex volume, out of which the convergence is wrong and fast, and inside of which the $V_G$ part leads to homogeneously fast convergence, while the $V_R$ contains guesses that lead to slow convergence (to a wrong solution).

\begin{figure*}[h]
   \centering
        (a)\includegraphics [width= 0.45 \textwidth] {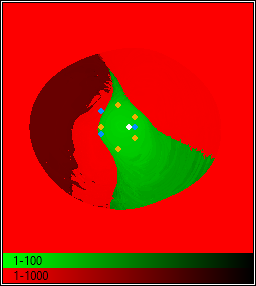}
        (b)\includegraphics [width= 0.45 \textwidth] {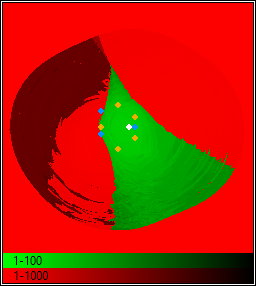}
        (c)\includegraphics [width= 0.45 \textwidth] {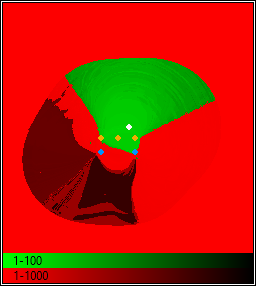}  
        (d)\includegraphics [width= 0.45 \textwidth] {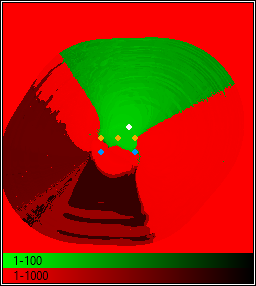}         
        
        \caption{2D convergence maps of $xy$ (a,b) and $xz$ (c,d) sections. The remaining 7 parameters are set to their exact values (a,c) or to their default values (b,d). The target (white dot) is in a more peripheral position (larger distance from the array axis), with respect to the case of Figure~\ref{fig:2dmapcentre}, at $z=31$mm. It is confirmed that using wrong guesses for $\vec B_g$ and $\vec m$ and wrong (larger) $z$ makes the ($x,y$) convergence area larger than using the exact values. }
  \label{fig:2dmapperif}
\end{figure*}
The Figure~\ref{fig:2dmapperif} represents convergence maps obtained when the dipole is located more peripherally, in $\vec r_p$. Again $xy$ and $xz$ sections are shown. They are obtained by assigning the fixed 7 parameters either their correct values or default guesses (details provided in the caption). Now, the shapes have more variegated aspects; however, also in this case, guesses with nearly axial $(x,y)$ coordinates lead systematically to good convergence, provided that $z$ is assigned in the correct interval, above $z_2$ and not too far from the sensors.

\begin{figure*}[h]
   \centering
        (a)\includegraphics [width= 0.45 \textwidth] {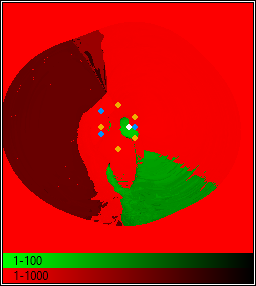}
        (b)\includegraphics [width= 0.45 \textwidth] {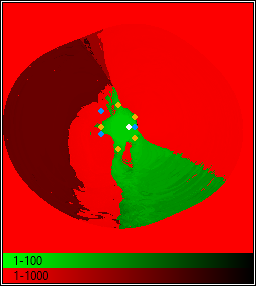}
        (c)\includegraphics [width= 0.45 \textwidth] {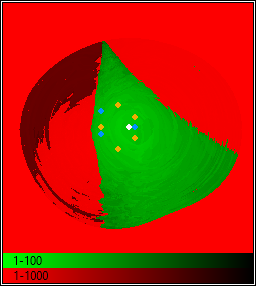}  
        (d)\includegraphics [width= 0.45 \textwidth] {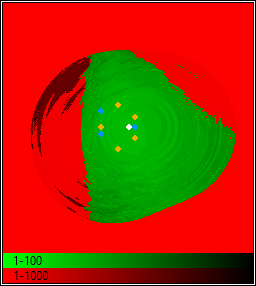}         
        \caption{2D convergence maps of $xy$ sections. The $\vec B_g$ and $\vec m$ guess is set at the default, and the $z$ guess is variously assigned: from 8~mm (a), to 20~mm (b), 60~mm (c) and 80~mm (d). The $z=$40~mm case was already shown in the Figure~\ref{fig:2dmapperif}-b. }
  \label{fig:2dmapzvaried}
\end{figure*}
A hint for an appropriate selection of $z$, is suggested by the maps represented in Figure~\ref{fig:2dmapzvaried}. This figure represents a $xy$ convergence maps obtained when the dipole is located peripherally (the same as for Figure~\ref{fig:2dmapperif}), the $\vec B_g$ and $\vec m$ guesses are assigned  their  default values, while $z$ co-ordinate is varied (details provided in the caption). As Figs.\ref{fig:2dmapcentre} c-d and \ref{fig:2dmapperif} c-d suggest, excessive $z$ values may lead to bad convergence, however intermediate values (e.g. z=60~mm to 80~mm) produce --with a larger number of iterations-- a wider convergence area in the $xy$ section. In contrast, wrong $z$ (Figure~\ref{fig:2dmapzvaried}-a) and small $z$ (Figure~\ref{fig:2dmapzvaried}-b) may reduce dramatically the convergence region.

Despite the complexity of the possible scenarios (intensity and orientation of the dipole and of the environmental field, accessible dipole positions with respect to the sensors, etc.) some general hints can be derived from the analysis of results shown in this section and in Sec.\ref{sec:trajectory}, \ref{sec:timeperformance}, which can be summarized as follows:

\begin{itemize}
    \item the initial guess of $\vec m$ is not critical (just assign reasonable values);
    \item the initial guess of $\vec B_g$ is not critical (just assign reasonable values);
    \item the initial guess of $\vec r$ is  critical, but whenever it is known that the target is on one side of the array (a given sign of $z$), and is not too displaced from the array axis, an axial guess for $(x,y)$ together with a reasonable guess for $z$ will work;
    \item the larger is the $z$ guess, the less critical is the  $(x,y)$ choice, but large values make the convergence slower, and too large values will prevent convergence;
    \item in case of correct guess, the iteration number depends weakly of the termination conditions and on the selected guess;
    \item good guesses bring to convergence within few tens of iterations;
    \item bad guesses with too large $|r|$  bring to (wrong) convergence within a few iteration steps;
    \item some wrong guesses with more reasonable $|r|$  bring to (wrong) convergence quite slowly, with many iteration steps (setting a limited number of iteration steps will help avoid wasting time with useless calculations);
    \item wrong convergence is easily detected, because the (local) minimum found is orders of magnitude larger than the absolute one;
    \item in case of a wrong convergence detected, trying other initial guesses having different $\vec r$ will help;
    \item in the considered application of eye-tracking, the limited RoI size makes a certainly-good guess possible;
    \item the  accuracy in reconstructing trajectories is generally good, but the best performance is obtained when the target moves on a surface nearly parallel to $\vec m$: in eye-tracking application, a radial orientation of $\vec m$ will be a favorite choice.
\end{itemize}

\section{Conclusion}
We have tested a new concept of an eye-motion tracker 
based upon a 8-sensor magnetic tracker and a small magnet. The performance in terms of accuracy, precision and speed has been analyzed, under different operating conditions. In particular, we have empirically studied the criticality (in terms of correctness and speed of the convergence) of assigning an appropriate starting guess to the numerical algorithm that infers the tracking parameters from the magnetometric measurements. The feasibility of a reliable sub-millimetric, 100Sa/s real-time 
tracking within a volume large enough to contain the RoI of eye-tracking experiments has been demonstrated.

The new technique described for eye movements recording seems have the possibility to reach a spatial and temporal resolution close to the magnetic search coils system in a portable and inexpensive device. Furthermore, the system will have the ability to record also head movements and 3D eye movements without a bulky and strongly tightened head mask. The major limit of the presented device may be associated with the comfort of wearing a contact lens with a millimetric magnet embedded. 
Eye movement recording and tracking has many important medical applications in diagnosis of neurological and vestibular disorders as well as in improving communication in severe neurological disorders.
This highlights the importance of developing a tracking instrumentation that is reliable, inexpensive, fast and accurate.

\section*{Patents} 
The accuracy and speed analyses at the focus of this work constitute a performance assessment of the hardware described in Ref.\cite{biancalana_instr_21}. A patent \cite{brevetto} is pending about inventions related to this research.

\bibliography{bibtrack}

\end{document}